# Gender Stereotypes Make Women Invisible: The Presence of Female Scientists in the Media

**Maider Eizmendi-Iraola[1] and Simón Peña-Fernández[2]***

1. University of the Basque Country (UPV/EHU) (ROR: 000xsnr85) maider.eizmendi@ehu.eus / https://orcid.org/0000-0002-5894-8238
2. University of the Basque Country (UPV/EHU) (ROR: 000xsnr85) simon.pena@ehu.eus / https://orcid.org/0000-0003-2080-3241

* Corresponding author

**Abstract**: Interest in the situation of women scientists has increased in recent years. Scientific institutions and governmental authorities have launched initiatives to promote science as a vocation among women. Within this context, the effort made by the media in disseminating role models is of vital importance. The aim of this research is to analyse the presence of women scientists in the media, and to conduct an appraisal of the status granted to them and the image conveyed about them. To this end, we analysed the content published on the websites of the regional newspapers Berria, Gara, El Correo and Diario de Navarra between 2014 and 2019 citing women scientists and researchers (N = 2362). The results indicate that presence is greater in quantitative terms, as is the prominence given to them in headlines. However, the disciplines in which they practise and the stereotypes which are conveyed shape their image, which in the case of STEM areas emphasise their role as carers and their singularity in the scientific field.

## 1. Introduction

Concern as to the situation of women in science emerged in the 1960s, when women entered both university studies and research centres. This is precisely the era of the study by the sociologist Alice Rossi, Women in science: why so few? (Rossi 1965). Since then, numerous initiatives, both scientific and institutional, have focused on the situation of women in the scientific field, in particular emphasising what are known as STEM subjects (Science, Technology, Engineering and Mathematics). In its resolution passed in December 2015, then, the United Nations approved the declaration of the International Day of Women and Girls in Science, which was held for the first time on 11 February 2016.

Along the same lines, equality between men and women is one of the axes of the European Commission's Horizon 2030 program and one of the main lines of the European Research Area. In Spain, for its part, Organic Law 3/2007 for the Effective Equality of Women and Men and Law

This is the author-accepted version of an article published in *Social Sciences*. The final version is available at: https://doi.org/10.3390/socsci12010030

14/2011 on Science, Technology and Innovation legally guarantee respect for the principle of gender equality in scientific field. In addition, following this legal framework, programs and activities have been developed in order to promote, among other aspects, the scientific vocations of women and girls. An example of this is the STEAM Alliance (Science, Technology, Engineering, Arts and Mathematics), an initiative of the Spanish Ministry of Education and Professional Training whose objective is to promote vocations in girls and young people and reduce the gender gap.

Women today represent 29.3% of research staff worldwide (UNESCO 2019), although the figures vary by country. In Spain, the average number of female scientists is 41%, although this percentage fluctuates significantly depending on the knowledge area or sector in which they are employed. Women are involved to a lesser degree in private enterprise and are under-represented in technological areas (She Figures 2019). Aside from these differences, in the scientific sphere women face difficulties in gaining promotion, and a greater proportion of women than men abandon their academic career. They thus face the “glass ceiling”, and suffer what is known as the “leaking pipeline” (Hunt 2016). Furthermore, in terms of output, they tend to publish less and receive fewer citations than men (Mueller et al. 2016; Duch et al. 2012), a pattern which is even repeated on those audio–visual platforms used for the dissemination of science (Eizmendi and Peña-Fernández 2021).

Research focused on the inequality faced by women in the scientific sphere has suggested such causes as difficulties in balancing academic and research work with caring responsibilities (Myers et al. 2020), mainly attributed to women; the prevailing androcentric values in science (Bird 2011), or the lack of valid role models encouraging scientific vocations among girls, and the existence of stereotypes (Carli et al. 2016). In general, girls and boys are socialized with different values and expectations for the future. Thus, for many factors that influence when deciding their professional career (Archer et al. 2020), the most prominent are those related to education that takes place both at school and in the family, as well as in the media. Precisely, the latter are today the priority sources when it comes to information about science (Mitchell and McKinnon 2019). In connection with this last aspect, and taking into account the task of the media in conveying gender roles, various studies have addressed the message conveyed as to women scientists. These studies have focused mainly on their function of building vocations, with recent research projects confirming that young people today still mainly associate scientific activities with men (Miller et al. 2018). They conclude that perceptions and expectations as to science professionals are more likely to be formed through exposure to representations of female scientists in the media than through direct contact with them (Chambers and Thompson 2020).

Vázquez-Cupeiro (2015) underlines that the representations of scientists in the media influence attitudes and behaviours and limit the vision of possible selves that, in turn, mark the training and professional itineraries. In this regard, although women have historically participated in the creation of knowledge, and despite the fact that the media have an important role in making women and their work visible (Chambers and Thompson 2020) in order to break with the gender gap that exists in the scientific and technological sector (Mena-Young 2018), their public recognition and their presence in the media (Francescutti 2018; García-Nieto 2015; Mena-Young 2018) and in fiction (Steinke 2017; Chambers 2022) is significantly smaller. Furthermore, beyond their quantitative presence, research confirms that they are treated differently from their male colleagues, in accordance with gender roles.

Thus, in texts about women scientists, the presence of gender stereotypes is observed, in a more implicit or explicit way (Miller et al. 2015). Through these powerful mechanisms, it is included in behaviours, interests and self-perception (Luong et al. 2019). The very earliest studies conducted in this sphere emphasised the stereotyped image associated with women scientists. In her research into the image of science professionals, Nelkin (1995) concluded that in the case of women there were frequent references to their femininity, while Shachar (2000) corroborated that in the case of women, the family role is used with regard to women to lend a different quality to the narrative. More recent studies also highlight the presence of gender stereotypes associated with the image (Mitchell and McKinnon 2019; Attenborough 2011; Chimba and Kitzinger 2010) and the role as carers (Husu and Tainio 2016; Tenglerová 2014) of women engaged in science.

As for character traits and qualities associated with science, media content tends to attribute to scientific performance those attributes traditionally assigned to men. Leadership and analytical capacity are associated with masculine roles (Carli et al. 2016), these being qualities more highly valued than those attributed to women, more connected with the emotional side (Husu and Tainio 2016).

Aside from the roles and stereotypes associated as a general rule with women, over recent years, messages about female scientists reveal another type of approach connected with their singularity, and associated with situations of inequality. Through such ideas, men continue to be represented as the norm, and women as the exception (Chimba and Kitzinger 2010; Niemi and Pitkänen 2017). Furthermore, female researchers who have proved successful, are seen to have an attitude which goes against gender norms. In short, women are thus represented as people acting in a predominantly masculine sphere (Husu and Tainio 2016).

Within this context, and bearing in mind the increasing number of initiatives aiming to promote the presence of women in science, this article aims to analyse the evolution of their visibility, and the main characteristics of the message conveyed about them, so as to ascertain whether there has been a change in the treatment afforded to them. The research questions raised are thus as follows:

RQ1. What is the media profile of female scientists and researchers?
RQ2. Which scientific disciplines attract greatest interest?
RQ3. What stereotypes and ideas are attributed to these female professionals?
RQ4. What are the characteristics of the journalistic texts in which female scientists are featured?
RQ5. Is there a relationship between the authorship of the content and the treatment given?

## 2. Materials and Methods

### *2.1. Selection of the Sample*

To answer these questions, this study focuses on an analysis of the content published on the websites of four regional newspapers in the Basque Country and Navarre -Gara, Berria, El Correo and Diario de Navarra- featuring female scientists and researchers. In addition to being four of the media with the largest number of readers in these territories, they have specific geographical, social and linguistic characteristics. El Correo belongs to the Vocento communication group. Their territorial reference is the province of Bizkaia and they publish content almost exclusively in Spanish. According to the data offered by Similarweb, El Correo receives 24.4 million monthly visits. Diario de Navarra, a newspaper published by Grupo La Información, focuses its activity in Navarra. Fully published in Spanish, their average number of monthly visits is 6.4 million, according to Similarweb. For their part, both Gara and Berria take Euskal Herria as a territorial reference, which includes the Autonomous Communities of the Basque Country and Navarre in Spain and the Northern Basque Country in France. Gara belongs to Baigorri Argitaletxea and publishes content in both Spanish and Basque, with an average of 2.1 million visits per month. Berria, on the other hand, is the only newspaper published entirely in Basque and has 1.2 million monthly readers.

The content analysis technique was employed for the research. This method, used repeatedly in media research (Agars 2004) has been strengthened in recent years thanks to the greater quantity of textual information available, mainly via the Internet (Andréu 2002). It is, then, an appropriate technique for gender analysis (Neuendorf 2011) and has frequently been applied in studying the image presented by the media of minority groups or those of particular interest (Wimmer and Dominick 1996).

In this case, a two-phase methodology was set out: first, an analysis of the general characteristics of all texts mentioning female scientists; and then, an exploration of the characteristics of the content, and the main ideas conveyed in content naming female scientists among the headline elements.

Beyond providing a current snapshot, the research aims to establish the evolution seen over recent years. The period considered thus runs from 2014 to 2019, a timeframe allowing us to study the impact of the growing attention that administrative and scientific institutions are giving two female scientists, which was intensified following on from the UN decision to declare 11 February to be International Day of Women and Girls in Science. The event was held for the first time in 2016, and the research period thus covers the two years before the three years after this.

As already stated, the sample comprises content citing female scientists and published in the media outlets selected for the study over the period 2014–2019. This thus required a search through a digital archive, using digital press libraries for the search as well as the websites' own search engines.

*2.2. Coding*

The search was performed by using keywords, as in other previous research studies (Segado-Boj et al. 2018; García-Nieto 2015). To specify the most relevant terms, a prior analysis was conducted of the publications in the media comprising the sample, along with consultation of previous studies. The decision was ultimately taken to employ the terms "investigadora" ("female researcher") and "científica" ("female scientist"). A subsequent screening process was nonetheless required.

It was also absolutely essential to clearly define the concept of woman scientist. Some research conducted in this field has employed a narrow focus, addressing only the field of STEM, and basing the study on the content published in scientific sections. In this case, the aim is to analyse female scientists and researchers engaged in all areas of knowledge, thus requiring a definition of the concept as those women engaged in academic and/or scientific activities, even if the journalistic content is not connected with such activities. We thus considered: (a) researchers and members of research groups publishing research results in any discipline, or those recognised for their efforts; (b) women expressing an opinion about a topic in the capacity of experts, and whose involvement is based on their scientific/academic activities or who, because of their leading status, take part at scientific activities and seminars, or have received awards; and (c) female scientists from history mentioned and/or highlighted for their scientific work.

In connection with the registration of the search unit and content research, Berria in its entirety, and Gara in part, publish their content in the Basque language. In these two cases the search was conducted by means of the terms "ikertzaile" and "zientzialari" (in the case of Gara, the search was performed in both Basque and Spanish).

The screening process was fundamental. The search engines in fact served to identify and inspect a total of 14,523 content elements, of which 2362 completed the sample in the first phase. Excluded from the sample were (1) the contents that exclusively cite male scientists; (2) those that employ a scientist and researcher, but without mentioning a specific person, for example, scientific activity; (3) those that are not related to scientific activity; and (4) the contents that are repeated. Therefore, the sieve makes it possible to define a sample that is more in line with the objectives set by the research.

Following definition of the sample, the analysis units, in other words the journalistic content published in the selected media outlets and referring to female scientists, were codified. In this case, the codification process was performed over two phases. In an initial phase, all the analysis units were studied, registering those variables connected with content characteristics (date, medium, authorship, section, purpose, area of knowledge). Subsequently, in a second phase a specific analysis was conducted of content in which female scientists appear in some element of the title. This entailed a new screening process, to separate out a total of 408 analysis units.

This entailed a new screening process, to separate out 408 analysis units. In this case, aside from the general details and variables, the analysis units were traced and coded to analyse. First, resources connected with scientific content are employed (technical words, allusions to the methodology...) and second, data they provide about these professionals (research area, position, geographical origin, adjectives used to describe them, and the data of context offered). In the latter case, the theme of the data provided has been analysed and it has been evaluated whether they refer to ideas that have traditionally been associated with women scientists. These ideas have been grouped into five categories: (1) Family and caregiving (e.g., "Never married, because when you're a chemist who's really after something, there's not much time to date"); (2) Physical appearance and clothing (e.g., "Elena Labarquilla is a woman with elegant shapes, she wears intense blue on a gray Tuesday"); (3) Character traits and suitability for science (e.g., "It is the example of obsessive dedication to a problem, how a certain bacteriophage works, a biological object that she wanted to understand how it replicated, how it infected bacteria... Everything. For her there was no life outside the laboratory"); (4) The extraordinary or pioneering nature of her work (e.g., "She is the only Basque researcher who has received ERC grants, one of the most prestigious worldwide"); and (5) Difficulties associated with their gender (e.g., "I remember that when I went to college, in Physics studies, about 60% were men and 40% were women. The difference was less pronounced. The problem is that as one advances in the scientific career these proportions vary greatly and to the detriment of women"). Along with the latter, it has also been analysed whether the contents focus on issues related to women: maternity and caring, female health and the body, gender discrimination and violence against women, and discrimination in science.

The data analysis and exploitation employed the programs Excel and SPSS. They were used to organise and cross-reference the data, serving to display the main characteristics of the media content and distinguish the existing relationships among the different variables.

## 3. Results

The results of the study are set out on the basis of the two constituent phases: the analysis of the content making mention of a female scientist; and the analysis of those publications in which such female professionals are included in the headline.

### *3.1. General Results*

The analysis of the distribution of the sample in the first phase (N = 2362) indicates an upward trend in the number of articles citing female scientists and researchers. Although the evolution is not linear, then, an upward trend may be seen in the presence of such female professionals (Figure 1). We must nonetheless analyse the characteristics of the texts in which they are presented, in order to afford us a more accurate insight into the visibility they are given.

Figure 1. Evolution of the presence of female scientists in information pieces.

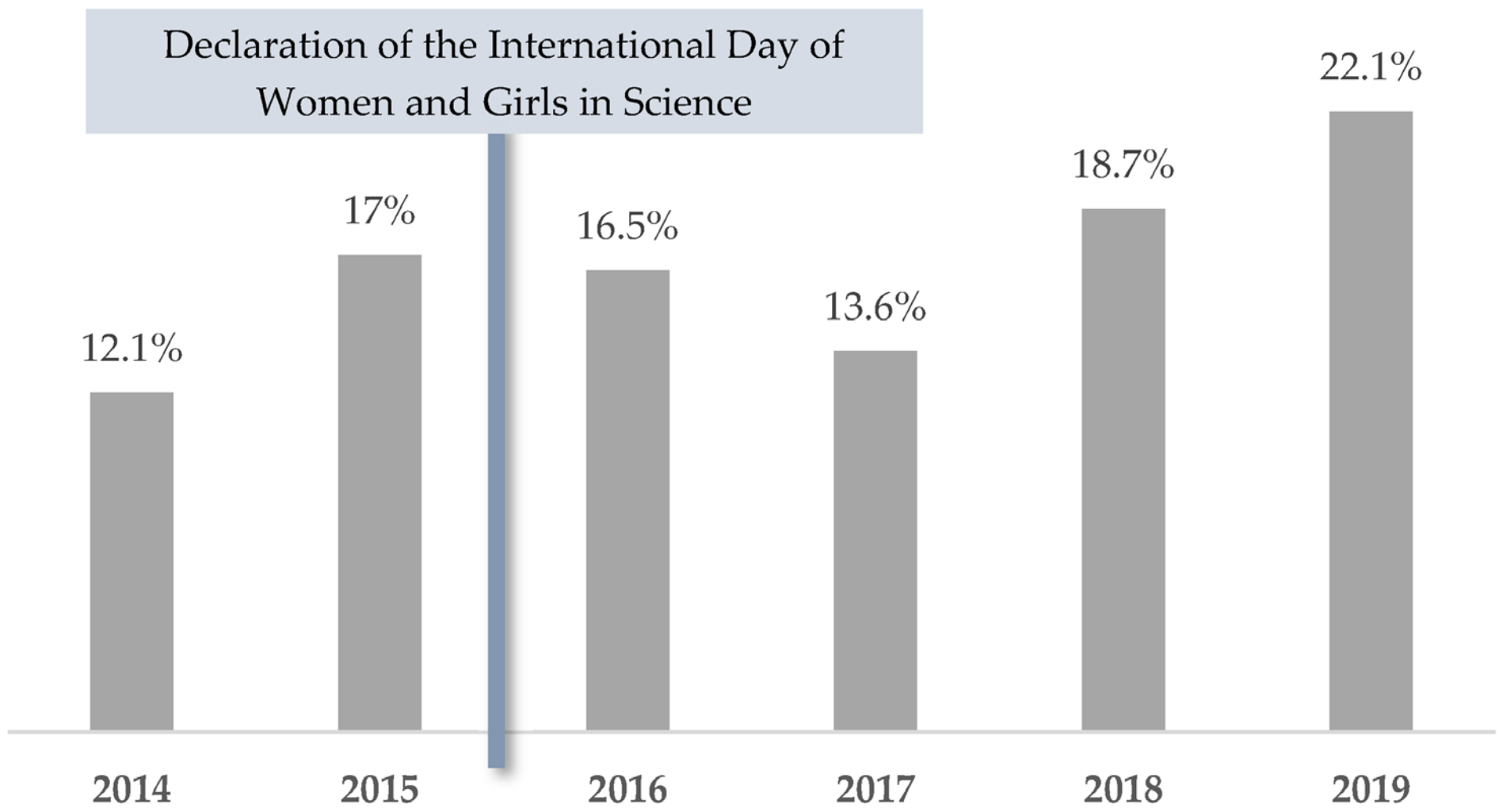

Source: Author's own work.

Despite the fact that research into female scientists has typically focused on STEM disciplines, an analysis of the topics of the publications indicates a significant presence of professionals in the areas of social sciences and the humanities, specifically 40.65%. History is the predominant discipline, with 7.86% of the texts concerning this field. It is followed by psychology (4.66%), economics (3.81%) and sociology (3.26%). As for natural and applied sciences (59.35%), health sciences and life sciences represent a significant proportion, with 21.29% and 14.23% content elements, respectively. Some considerable way behind are items concerning chemistry (4.99%), earth sciences and space (4.66%), and technology (4.28%).

Almost half of the texts analysed (46.27%) are for the purpose of explaining the results of a study or research project. A significantly lower percentage (21.59%) of the content is intended to examine and explore a specific topic in greater depth, publications in which the researchers are presented as sources, although in most cases there is no reference made to their academic career. The remaining items analysed correspond to the categories of scientific activities and seminars (17.32%) and awards and accolades (9.99%). Lastly, 3.81% of content refers to female scientists in circumstantial terms.

With regard to authorship (Figure 2), a notable proportion of the content is produced by information agencies (31.2%). This figure tallies with those presented by De Semir and Revuelta (2006), asserting that smaller media outlets often use these sources, given the absence of specialist science journalists on their teams. As for articles written by individual journalists, the figure is similar, although the percentage of men (21.55%) is slightly higher than for women (19.39%). However, a detailed examination of the contents addressed by gender reveals the existence of a clear difference in terms of the scientific areas covered. Except for health sciences and life sciences, which are the areas most often covered both by men and by women, in the case of the latter, the content most typically relates to the social sciences, above all anthropology, law, education, psychology, philosophy and communication sciences. In the case of male journalists, the main areas discussed are astronomy, physics, chemistry and space and earth sciences. The data indicate that the agenda of the journalists does not, perhaps, guarantee a greater presence, but does influence the areas of knowledge addressed by the media outlet.

Figure 2. (a) Authorship of content (b) Percentage of knowledge areas according to authorship.

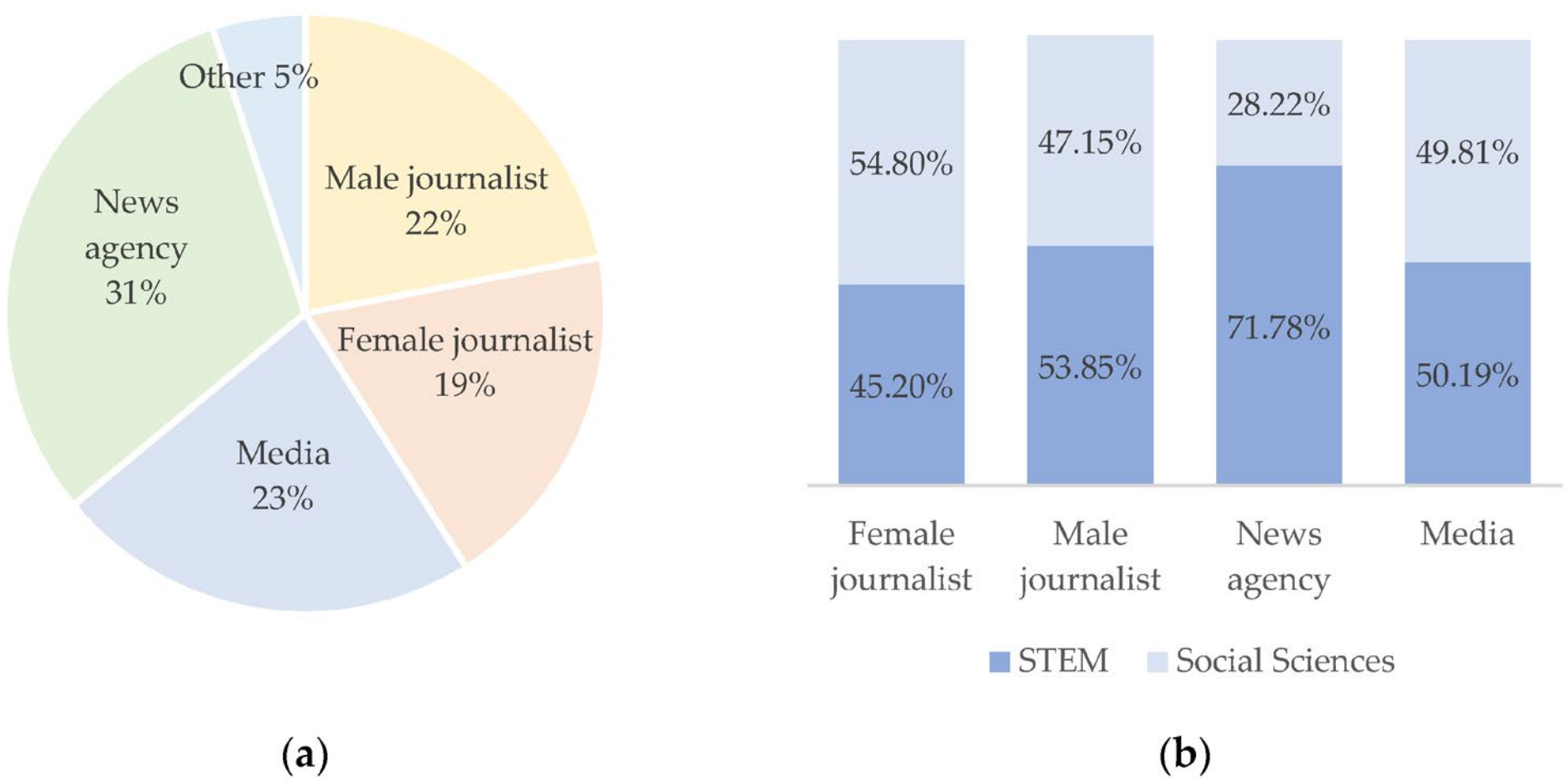

Source: Author’s own work.

As for the sections in which such content appears, we see a predomination of those areas including news from various topics, such as institutional policy, education, health… provided that the geographical context corresponds to the region covered by the information enterprise. In the media outlets analysed, these sections are designated as Biscay (El Correo), Navarre (Diario de Navarra) and Basque Country (Gara), and account for 21.34% of content. Society is also a recurrent section (18.05%), above all in the case of Berria. One could, then, state that content about female scientists corresponds more to those sections covering more varied topics, and within this context, supplements and features sections also reveal a considerable percentage (7.66%).

In the case of science, none of the media analysed had such a section during the period covered. Consideration was nonetheless given to the fact that topic headers are used to classify some content as scientific. A total of 266 news items were classified under this criterion, in other words 11.26%, the majority of which (67.40%) cover topics connected with the hard sciences, with a particular emphasis on health sciences and life sciences. The fact is that an analysis of content about women scientists likewise reveals a clear trend towards “medicalisation” (Dunwoody 2008).

Meanwhile, the culture section publishes 9.14% of the news items featuring women scientists, mainly content connected with the history and sciences of arts and literature.

*3.2. Female Scientists in Headlines*

A second phase analysed those texts in which women scientists and researchers appear in the elements that make up the headline (N = 408). In this case, a continuous progression may be noted during the research period (Table 1), seen above all in 2016 and 2017. It should be pointed out that in December 2015, the UN General Assembly declared 11 February to be International Day of Women and Girls in Science, and one may therefore reasonably imagine that this event had a positive impact on the presence and visibility of female scientists in the media analysed.

Table 1. Evolution of content addressing the inequality of women in science.

| | Women in the Title | Content Concerning Inequality | Percentage |
|---|---|---|---|
| 2014 | 32 | 1 | 3.12% |
| 2015 | 46 | 3 | 6.52% |
| 2016 | 56 | 9 | 16.07% |
| 2017 | 68 | 12 | 17.65% |
| 2018 | 94 | 12 | 12.77% |
| 2019 | 112 | 20 | 17.85% |

Source: Author's own work.

However, visibility should not be understood only in quantitative terms. We must analyse the role ascribed to these female professionals as scientific figures. It should in this regard be pointed out that despite being presented in the headline, in 18.87% of cases they are merely cited, without being given any voice, thereby obviating any role as authorised sources, and hence very much limiting their status as scientific figures.

Among the women scientists included in headlines, 29.41% are featured in content focusing on the awards or bursaries they have received (Figure 3). Public recognition of such women therefore represents an effective mechanism in terms of headline visibility. Unfortunately, women are poorly represented when compared with men in the field of scientific awards (López-Sancho 2010; Francescutti 2018). The remaining mentions are to present the results of a research project (30.14%), as a source for greater insight into a topic (23.28%), or as a participant or representative at a gathering (10.78%).

As for their value as sources of information, a significant role is here played by the details provided both as to their academic career and references to current or prior research activities. In this regard, we analysed the mentions made in the texts as to methodology, to be found in 47.79% of the units analysed. No major difference is found between social sciences and humanities in terms of the percentage of mentions of the research procedure or method, although the form does change. While for what are known as the hard sciences, it is the journalists themselves who allude to or ask about this, in the case of the social sciences, it is more often the female protagonist herself who mentions them. One may, then, perceive a more apparent need to proclaim the scientific nature of such areas, and to emphasise the method or procedures employed.

Figure 3. Distribution of the percentage of women in the title, according to the analysis units comprising the sample.

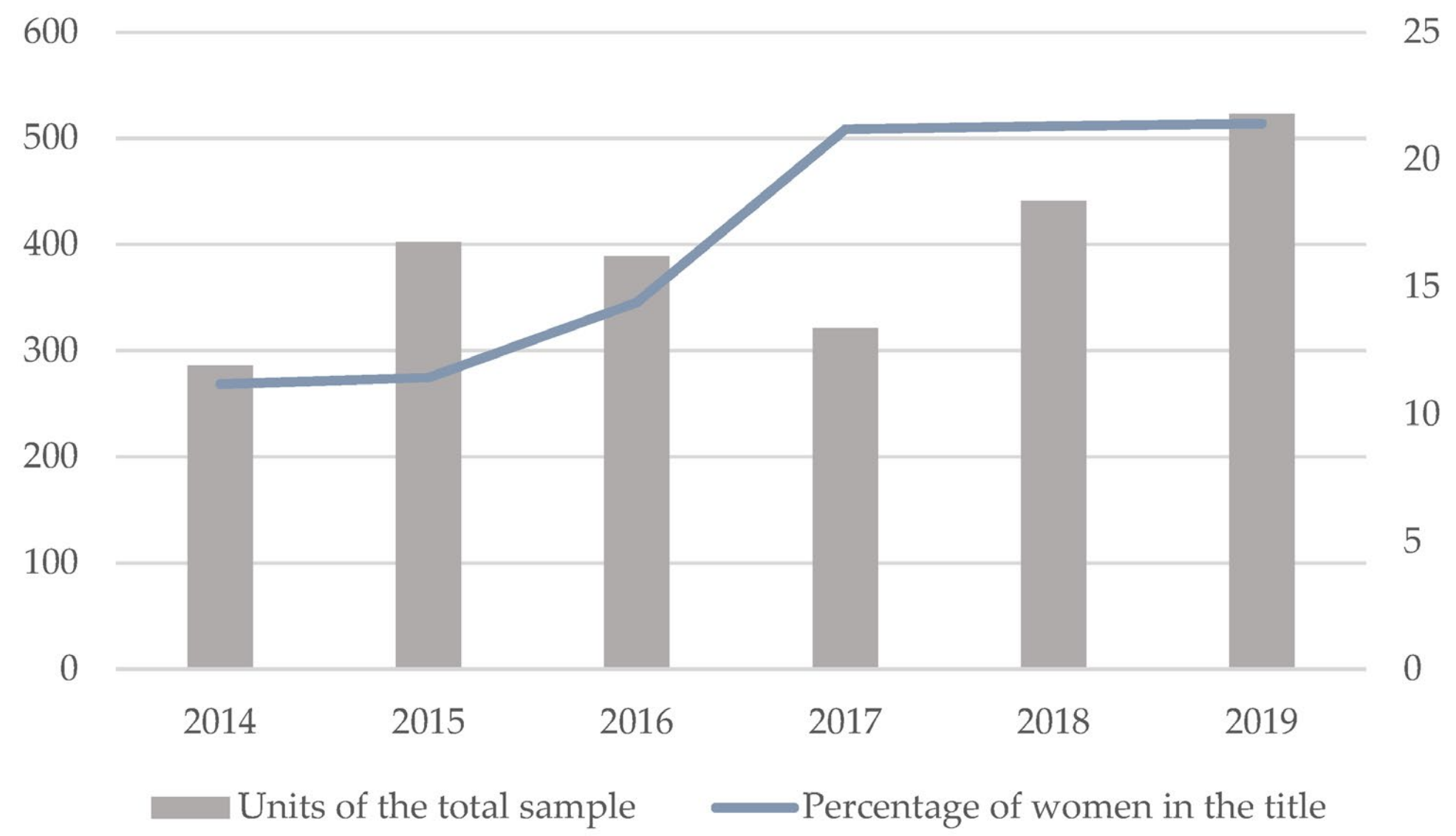

Source: Author’s own work.

*3.3. Stereotypes and Ideas*

The study analysed the ideas and stereotypes conveyed about women, classified in terms of five topics: Family and care; physical appearance and clothing; personality and scientific qualities; singularity; and gender-related difficulties. In this case, a clear difference may be seen among disciplines (Figure 4). As is apparent, the presence of stereotypes associated with women scientists is considerably higher in the case of the hard sciences.

Figure 4. Presence of stereotypes, by knowledge area.

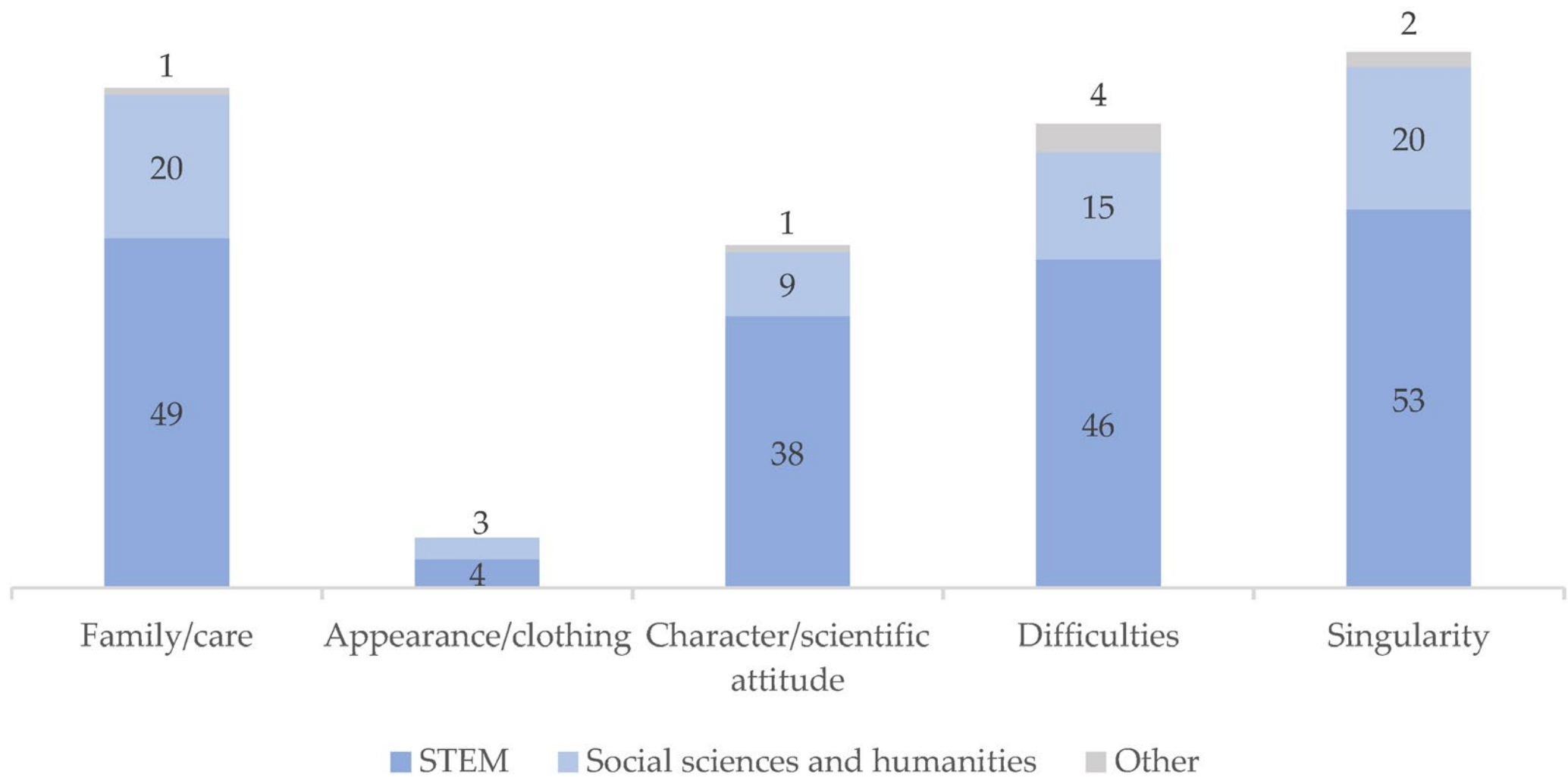

Source: Author’s own work.

As for the development of these ideas, the results show that stereotypes associated with the appearance and clothing of female scientists have declined considerably, if one takes into account studies such as that by Chimba and Kitzinger (2010), who asserted that this type of concept appeared in half of all content concerning women scientists. The same does not occur in the case of references to family and caring duties. They may be found in 17.15% of texts, formulated in two different ways: firstly, as biographical information; and secondly, to represent the clash between academic career and family. References to their immediate context, typically to men, likewise appear as an explanation for their scientific vocation.

There are notably references to difficulties and the singular status of such female professionals, present in 15.9% and 18.3% of the content analysed in this second phase, respectively. References to exceptional status are connected with the social mindset as to the role of scientist, mainly associated with men. Women engaged in science are therefore presented as symbols (Shachar 2000). Both the latter, and ideas associated with gender-based difficulties, are found above all in content regarding women engaged in the hard sciences, with a particular emphasis on chemistry and physics.

Combined with this last idea, we find that content focused on matters connected with women, such as maternity and caring, female health and the body, gender discrimination and violence against women, and discrimination in science, account for 26.22% of the publications analysed. With regard to this last issue, there has been a considerable increase (Table 1), mainly from 2016 onwards, the year when the International Day of Women and Girls in Science was first held. There is, then, a growing interest in addressing this issue.

The aforementioned stereotypes and concepts are underpinned by the adjectives employed to describe women scientists. These designations are classified into four groups:

(a) Those adjectives emphasising the success and unusual nature of their careers: noted, best, brilliant, multifaceted, precocious, talented, leading experts…
(b) Those terms defining their personality: good, sincere, passionate, committed, innovative, considered, contagiously joyful, with strength, strong, curious, dreamer, hard-working…
(c) Those referring to their attitude towards science: obsessive, indefatigable, tireless, persevering…
(d) Those highlighting the singularity of their role: rebel, unusual, maverick…

These adjectives are in line with the ideas conveyed about women scientists, as they highlight the woman’s excellence, underline her virtues, emphasise her persevering nature, and spotlight her singularity.

## 4. Discussion

The content analysed reveals an increase in the number of women scientists present in the media. They are also taking on a more prominent role, if one bears in mind their presence in the different elements of the title, which is increasing year on year. This increase coincides with the timing of the declaration of the International Day of Women and Girls in Science, approved in December 2015 by the UN General Assembly, and held for the first time in February 2016. The conclusion is thus that the event, instigated with the aim of championing equality for women in the field of science, has provided the media with a helpful context to include the matter of women scientists on their agenda: first of all, through content produced by editorial teams on their own initiative; and secondly, through coverage of the initiatives organised by different institutions and agents.
Within the context of visibility, however, we must specify details beyond the number of occasions on which women scientists are featured. The very concept must be understood in a broader sense, tied to the space and status afforded to female protagonists. Hence, the fact that, despite their increased presence, an analysis of the characteristics of such content reveals that in many cases these women scientists and researchers are only mentioned, and above all the space afforded to them is in regional news sections grouping together all manner of topics, and in society sections. They have little presence in the content which the media outlets themselves define as science, which is furthermore very much linked to areas classified as hard science.

In this regard, one of the contributions made by this study is an analysis of the participation of women scientists from a comprehensive perspective, taking into account all areas of knowledge, including the social sciences and humanities. Previous studies have focused their research on the hard sciences, more specifically what are known as STEM, based on a lack of reflection, definition and consistency as to what we mean by science (Cassidy 2021), and hence science journalism. The data analysed indicate that the percentage of women presented in the media and working in these two areas is 40.65%. Therefore, from a methodological point of view, it may be seen that this comprehensive perspective facilitates a more wide-ranging and more closely aligned analysis in studying the message conveyed as to such female professionals, since the area in which they work has a direct impact on the status afforded to them. This study in fact demonstrates the existence of two strategies which underpin a questioning of the role of women as creators of scientific knowledge: First of all, the classification of scientific disciplines themselves; and furthermore, the use of gender-related stereotypes.

Concerning the former aspect, although there is a significant presence in the sample of women researchers engaged in the humanities and social sciences, the way these disciplines are treated does not correspond to the parameters attributed to scientific journalism. Such professionals are in the main presented in sections covering social issues, and furthermore presented in content the purpose of which is to explore a topic in greater depth, and the media therefore typically put them forward as "interpreters" of society and social trends. As a result, although their selection is based on the legitimacy derived from their academic track record, they are cited without any reference to specific research or studies, and their contribution is thus expressed with the status of "opinion pieces".

In the case of women professionals engaged in other areas, the scientific nature of the disciplines or of those researching the field are essentially not questioned. The status of women researching in the field of the hard sciences is more dependent on the message conveyed, and above all the stereotypes and ideas it includes. The data have demonstrated that it is precisely in the stated fields of knowledge where we see the greatest conflict between the gender role and role of female scientist (Knobloch-Westerwick et al. 2013).

As for stereotypes, differences were detected with regard to previous studies: the most obvious stereotypes, concerning physical appearance and clothing, were scarcely registered, in proportional terms. The same cannot be said for references to caring roles and the family. The existing literature has extensively analysed the tendency to link professional women and references to the family, a tendency tied to the caring role traditionally attributed to women. In the case of women researchers, as the data from this study reveal, references to the family may be found in almost one in every five texts, both by way of biographical details and to illustrate the existing conflict between an academic career and the family (Mitchell and McKinnon 2019). Following on from this last point, one of the purposes of such references, on the part of both journalists and the featured women themselves, is to raise the issues of work–life balance which exist in the scientific field. In any event, the use of such references serves to position the narrative regarding women in science within a family context.

The aforementioned references to the family also reveal aspects connected with stimulating vocation. This tendency underpins the "idea of singularity" corresponding to women still being engaged in scientific work, in other words, women working in science and research necessarily require a specific and stimulating context leading them in this direction, often linked to male figures, such as husbands, brothers or teachers.

One of the conclusions of this research is precisely the power of messages positioned within the context of exceptionality and difficulty in the image presented of women scientists. An explicit analysis was also conducted of the references to these two ideas, and in fact the content presenting female scientists in the headline reveals the most commonly repeated idea to be how extraordinary their story is (18.38%), and the difficulties they face.

The narrative about women scientists has therefore revealed the continued importance given to their singularity in the scientific sphere, and the gender identity of female researchers. However, changes can be seen in how this idea is conveyed. While the studies conducted previously emphasised references to the physical appearance and clothing of women, this research identified to a greater extent the other messages which have come to the fore, linked to difficulties and exceptionality, and above all representing the women researchers featured in the media from the perspective of their female gender. Therefore, although there is a qualitative difference, the message currently conveyed with regard to women scientists is based on the same idea, representing science as a man's field, and hence tending to reflect women in terms of singularity and difficulty.

The danger involved in conveying such ideas lies in undermining the image of women as scientists (Steinke 2017), because the focus is placed not on their abilities and contributions, but rather their singular status and difficulties, and in short their position as women, which is seen as the root factor. This tendency was noted specifically in a number of the headlines analysed, using the word 'woman' both to refer to and to describe the featured individual. In many cases, the aim is specifically to highlight the unequal situation faced by female professionals, but the result impacts on their scientific status.

With regard to their legitimacy as researchers, another of the tendencies identified is to present women scientists in content connected with topics traditionally linked to femininity, and above all gender discrimination, or inequalities in the scientific sphere, this being noted in 26.22% of those texts featuring women in their title. Women scientists are thus often the key figures in topics in what is seen as "women's" issues. In short, following the reasoning of Sánchez-Calero et al. (2013, p. 14), such female scientists symbolically represent women, rather than professionals in the field of science. In many cases, then, they are presented essentially as sources of information as to the situation of women in science.

## 5. Limitations

This study has limitations in terms of the sample analysed, which takes the regional media of the Basque Country as a reference. In addition, it would be interesting to analyze the treatment that the media offer to male scientists and academics in order to determine if there are decisive quantitative and qualitative differences compared to women.

The results also suggest the need to continue investigating the messages transmitted by the media and delve into the factors that shape them, both through the analysis of the currently predominant structures in the scientific field, as well as a communicative perspective on the organization and media routines.

Likewise, the study can contribute to future research for the qualitative study on the collective and subjective factors that the researchers perceive, and that determine the participation of women scientists both in the media and in other dissemination activities.

## 6. Conclusions

Although the study confirms progress in the presence and visibility offered in the media as to women scientists and researchers, the results indicate that concern as to the inequality faced by women in the world of science has not been reflected in an approach focusing on their contributions and their status as scientific figures. Women working in the field of the social sciences and humanities are impacted by the lower social status that has historically been afforded to disciplines in these fields, which likewise means that women working in these branches are presented as “interpreters”, rather than scientific experts. In the case of STEM subjects, gender stereotypes and ideas connected with inequality are those which, by prioritising their feminine role, undermine the scientific role of these women.

Despite the advances that women have demonstrated both in the academic world and in scientific activities, then, science is still represented in the media as a male environment in which women represent the exception. This framework demonstrates the need to continue researching the message conveyed and to explore in greater depth the factors shaping this, both through analysis of the currently predominant structures on the scientific landscape, and a communicative focus on the organisation and routines of the media.